\documentclass[aps,pra,twocolumn,showpacs,groupedaddress]{revtex4}

\usepackage[dvips]{graphicx}
\usepackage{amsmath}
\usepackage{float}
\usepackage{bbm}
\newcommand{\half}{\mbox{$\textstyle \frac{1}{2}$}}
\newcommand{\ket}[1]{\left | \, #1 \right \rangle}
\newcommand{\kets}[1]{| \, #1 \rangle}
\newcommand{\bra}[1]{\left \langle #1 \, \right |}

\newcommand{\brakets}[2]{\langle\, #1\,|\,#2\,\rangle}

\newcommand{\av}[1]{\langle #1\rangle}

\newcommand{\vac}{\ket{\textrm{vac}}}

\newcommand{\eqr}[1]{Eq.~(\ref{#1})}
\newcommand{\figr}[1]{Fig.~\ref{#1}}
\newcommand{\secr}[1]{Sec.~\ref{#1}}
\newcommand{\appr}[1]{Appendix~\ref{#1}}

\def \ic {\textrm{i}}

\begin{document}

\title{Adiabatic melting of two-component Mott-insulator states}

\author{M. Rodr\'iguez$^{1,2}$, S. R. Clark$^1$ and D. Jaksch$^1$}
\affiliation{$^1$ Clarendon Laboratory, University of Oxford,
Parks Road, Oxford OX1 3PU, U.K.} \affiliation{$^2$ ICFO-Institut
de Ci\`encies Fot\`oniques, 08860 Castelldefels (Barcelona),
Spain}

\begin{abstract}
We analyze the outcome of a Mott insulator to superfluid
transition for a two-component Bose gas with two atoms per site in
an optical lattice in the limit of slow ramping down the lattice
potential. This manipulation of the initial Mott insulating state
transforms local correlations between hyperfine states of atom
pairs into multiparticle correlations extending over the whole
system. We show how to create macroscopic twin Fock states in this
way an that, in general, the obtained superfluid states are highly
depleted even for initial ground Mott insulator states.
\end{abstract} 

\maketitle

\section{Introduction}
The experimental realization of degenerate atomic gases \cite{BEC}
have opened up the possibility of engineering strongly correlated
many-body quantum states. Such states are fundamental for the
development of quantum technologies such as those implementing
quantum information protocols and quantum computation. A paramount
example is the realization of a Mott insulator (MI)
\cite{SFMI,SFMIT} in the lowest Bloch band of an optical lattice
which can serve as a quantum register \cite{spinlattice}.
Bose-Einstein condensates of alkali atoms have also been proposed
as systems to realize macroscopic superposition states \cite{GHZ}
or number correlated states (also called twin Fock states) which
are useful for quantum metrology \cite{Holland, Ketterle06,
Kheruntsyan, rodriguez}.

In \cite{rodriguez} we proposed a method for engineering twin Fock
states via a two-component MI to superfluid (SF) transition. The
proposed scheme manipulates hyperfine states of atom pairs pinned
to single lattice sites and decoupled from one another in the MI
regime \cite{spinlattice,twoatom}. The MI limit has already proven
to be useful starting point for quantum state engineering since
the motional dynamics are frozen out and the system decouples to a
set of single site Hamiltonians to good approximation. In the case
of a two-component system with doubly occupied sites spin-changing
collisions will occur and can be accurately controlled with
external magnetic or microwave fields \cite{twoatom}. As a result
the spin-dynamics of atom pairs behaves like an effective
two-level system undergoing Rabi-like oscillations. In contrast to
the usual Rabi-model for the coupling of a pair of {\em
single-particle} states this represents a coherent coupling
between a pair of {\em two-particle states} induced by collisional
interactions. These local {\em two-particle} correlations are then
transformed into {\em multi-particle} correlations extending over
the whole system by melting the MI into a SF. This quantum melting
can be experimentally implemented by adiabatically ramping down
the depth of the lattice potential.

The build up of long-range correlations when a periodic potential
is dynamically lowered is a highly non trivial process that occurs
rather quickly in a one-component system
\cite{SFMI,rodriguez,Zurek}. The quantum melting of a
multi-component system is still an open problem that is even more
intriguing due to the different phases that appear in the $T=0$
phase diagram depending on the ratios of the interparticle
interactions and of the tunneling rates of each component
\cite{Demlerymas}. An analysis of the two-component symmetric case
with all interparticle interactions equal can be found in
\cite{Ripoll}. In \cite{rodriguez} we studied the adiabatic
melting of an optical lattice in a two-component system with
different values for the inter-species and intra-species
interactions. We showed that one can create twin Fock states with
high fidelity via melting of an initial MI state with two atoms
per site in different internal states. We inspected possible
sources of experimental imperfections such as atom loss during
melting or depletion caused by final finite interactions. The
possibility of creating macroscopic superposition states via the
adiabatic melting of on-site entangled MI states in an infinitely
connected lattice was also briefly considered in \cite{rodriguez}.
An examination of the adiabatic melting in an infinitely connected
lattice, including the effect of the dynamical relative phase
acquired by the state can be found in \cite{conf}.  In this paper
we extend the results in \cite{rodriguez,conf} and analyze the
melting of on-site superposition states in a lattice with only
nearest neighbor hopping. We show that the final SF states
obtained from such on-site superposition states are highly
depleted even if the initial state is a ground MI state. For the
melted states we calculate the momentum distribution and the
momentum space particle number correlations. These quantities are
experimentally accessible either by time of flight expansion of
the cloud or via shot noise measurements \cite{EX}.

This paper is organized as follows. In section \ref{sec:bhm} we
introduce the two-mode Bose-Hubbard model (BHM) and the initial
Mott insulating states that we consider. In section
\ref{sec:adiabatic} we describe the adiabatic approximation and
the no-crossing rule that we use to infer the final superfluid
states. In section \ref{sec:melt_ab} we illustrate how one can
create macroscopic twin Fock states if the inter-species
interaction is greater than the intra-species interaction during
melting. In \ref{sec:messy} we show using the symmetry properties
of the initial states and a series of approximations, that an
on-site entangled MI ground state melts into a highly depleted SF
state in the limit of intra-species interaction greater than the
inter-species interaction.


\section{\label{sec:bhm}Two-component BHM}
In this section we introduce the two-component BHM that describes
the dynamics of ultra-cold atoms in an optical lattice. We discuss
its experimental feasibility and analyze the symmetries of the
Hamiltonian. Finally we introduce the initial MI states under
consideration.
\subsection{Optical lattice setup}
 Our focus is on a state-dependent
optical lattice which is sufficiently deep that two hyperfine
states $a$ and $b$ of the atoms are be trapped in the lowest Bloch
band of the lattice. The resulting system is then accurately
described~\cite{SFMI} by the two-component BHM with the
corresponding Hamiltonian ($\hbar=1$)
\begin{eqnarray}
&&\hat{H}=-\sum_{\langle i,j \rangle}(J_a \hat{a}^\dagger_i
\hat{a}_j + J_b\hat{b}^\dagger_i\hat{b}_j) + U \sum_i
\hat{n}_i^a\hat{n}_i^b
\nonumber \\
&&+\frac{V_a}{2} \sum_{i} \hat{n}_i^a(\hat{n}_i^a-1)+\frac{V_b}{2}
\sum_{i} \hat{n}_i^b(\hat{n}_i^b-1),\label{eq:ham}
\end{eqnarray}
where $\hat{a}_{i}(\hat{b}_{i})$ is the bosonic destruction
operator for an $a(b)$-atom localized in lattice site  $i$,
$\hat{n}_i^a=\hat{a}^{\dagger}_i\hat{a}_i$ and
$\hat{n}_i^b=\hat{b}^{\dagger}_i\hat{b}_i$, while $\langle i,j
\rangle$ denotes summation over nearest-neighbors. The parameter
$J_{a(b)}$ is the tunnelling matrix element for atoms in state
$a$($b$); $V_{a(b)}$ and $U$ are the on-site intra- and
inter-species interaction matrix elements, respectively. A crucial
feature of the optical lattice setup is that the ratio between the
tunnelling and interaction matrix elements in \eqr{eq:ham} is
experimentally controllable via the lattice laser intensities.
Further independent control of the interaction matrix elements
$U$, $V$ can be achieved with Feshbach resonances~\cite{Feshbach},
or by shifting the $a$ and $b$ atoms away from each other using
state-dependent lattices~\cite{spinlattice}. All of these methods
of control can be exploited dynamically to permit different
regimes of the Hamiltonian in \eqr{eq:ham} to be unitarily
connected. Together with the long decoherence times of ultra-cold
atoms this dynamical manipulation can often be implemented on
near-adiabatic timescales.

Based on the considerable flexibility of the optical lattice setup
we make a number of simplifications in our analysis. First, we
restrict the couplings to the symmetric case where $J_a=J_b=J$ and
$V_a=V_b=V$. Second, we consider states of the system $\ket{\psi}$
in which the lattice is commensurately filled with two atoms per
site so $\hat{N}\ket{\psi} = N\ket{\psi}$ has $N=2M$, where
$\hat{N} = \sum_i(\hat{n}_i^a + \hat{n}_i^b)$ is the total
particle number and $M$ is the total number of sites. Finally, we
confine our attention to a 1D lattice with periodic boundary
conditions (i.e. $M+1 \equiv 1$) in addition to the spatial
homogeneity of the matrix elements in \eqr{eq:ham}. This is not
only a good approximation for the center of a large system, where
trapping and boundary effects have a negligible influence, but is
also directly applicable to recently demonstrated ring-shaped
optical lattices~\cite{Cataliotti}.

\subsection{Symmetries}
\begin{figure}[t]
\includegraphics[width=8cm]{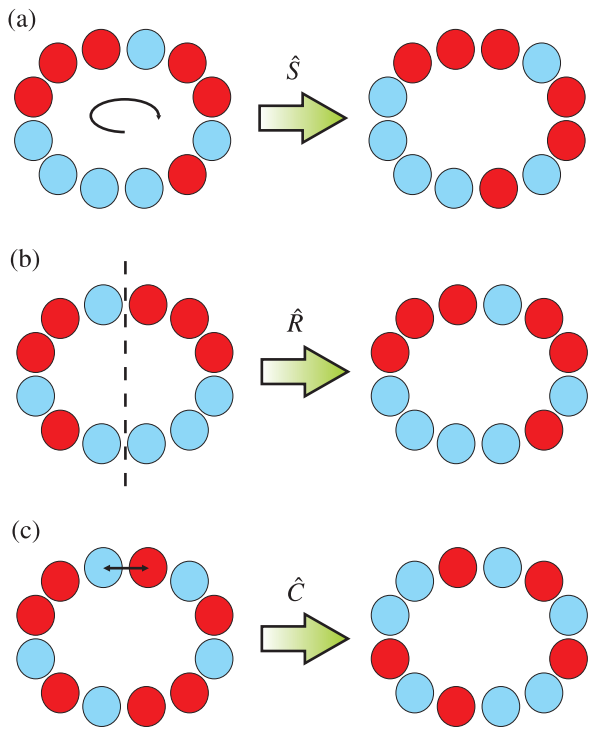}
\caption{Schematic representations of (a) cyclic shifting, (b)
mirror reflection and (c) color inversion symmetry for a 1D
lattice with periodic boundary conditions. The colors represent
atoms in two different hyperfine states $a$ and $b$.}
\label{fig:sym}
\end{figure}

The most basic conserved quantity of the two-component BHM is the
total number of atoms $\hat{N}_{a}= \sum_i \hat{n}^{a}_i$,
$\hat{N}_{b}= \sum_i \hat{n}^{b}_i$ in the either of the internal
states $a$ or $b$, respectively. With the additional assumptions
outlined a number of other symmetries arise which we exploit in
this work. The homogeneity and periodic boundary conditions
assumed results in translational invariance. To exploit this we
introduce quasi-momentum operators for each component as
\begin{eqnarray}
  \hat{c}^{\dagger}_k &=&
  \frac{1}{\sqrt{M}}\sum_{i=1}^M\hat{a}^{\dagger}_i e^{-\ic k r_i}, \nonumber \\
  \hat{d}^{\dagger}_k &=&
  \frac{1}{\sqrt{M}}\sum_{i=1}^M\hat{b}^{\dagger}_i e^{-\ic k r_i}. \nonumber
\end{eqnarray}
where $k$ is the quasi-momentum in the first Brillouin zone and
$r_i$ is the position of the site $i$. The quasi-momentum states
spanning the first Brillouin zone have $k=\delta k \times
\{-K_m,..,0,..,K \}$ with $\delta k = 2\pi/Md$ and $d$ being the
distance between lattice sites. For $M$ even $K_m=M/2$ and
$K=M/2-1$, while for $M$ odd $K_m=K=(M-1)/2$. The total momentum
of the system is then $\hat{P} = \sum_k k\,(\hat{{\tt n}}^a_k +
\hat{{\tt n}}^b_k)$, where $\hat{{\tt n}}^a_k =
\hat{c}^{\dagger}_k\hat{c}_k$ and $\hat{{\tt n}}^b_k =
\hat{d}^{\dagger}_k\hat{d}_k$, and is a conserved quantity. The
total momentum generates {\sl cyclic} shifts of the lattice sites,
e.g. $i \mapsto i+1$ for the site indices of all operators, via
the unitary $\hat{S} = \exp(-2\,\ic \hat{P} \pi/M)$. Its action on
the bosonic creation operators is
\begin{eqnarray}
  \hat{S} \,\hat{a}^{\dagger}_i\, \hat{S}^{\dagger} &=&
  \hat{a}^{\dagger}_{i+1}, \nonumber \\
  \hat{S} \,\hat{b}^{\dagger}_i\, \hat{S}^{\dagger} &=&
  \hat{b}^{\dagger}_{i+1}, \nonumber
\end{eqnarray}
simultaneously for all sites $i$. The homogeneity of the system
also gives mirror reflection symmetry in which site indices for
operators interchange as $i \leftrightarrow M-i+1$. In this case
the {\sl parity} operator $\hat{R}$ is the conserved quantity and
implements the transformation
\begin{eqnarray}
  \hat{R} \,\hat{a}^{\dagger}_i\, \hat{R} &=&
  \hat{a}^{\dagger}_{M-i+1}, \nonumber \\
  \hat{R} \,\hat{b}^{\dagger}_i\, \hat{R} &=&
  \hat{b}^{\dagger}_{M-i+1}, \nonumber
\end{eqnarray}
simultaneously for all sites $i$. Finally, the homogeneity and
symmetry of the matrix elements result in {\sl color} inversion
symmetry $a \leftrightarrow b$. Like parity there is a
hermitian-unitary operator $\hat{C}$ which is the conserved
quantity and implements the transformation
\begin{eqnarray}
  \hat{C} \,\hat{a}^{\dagger}_i\, \hat{C} &=&
  \hat{b}^{\dagger}_{i} \nonumber,
\end{eqnarray}
simultaneously for all sites $i$. These three symmetries are
schematically represented in \figr{fig:sym}(a)-(c).

Symmetries are by definition related to the presence of
degeneracies~\cite{Nussinov} or more precisely to a degeneracy
which remains for all values of the couplings in the Hamiltonian -
a so-called permanent degeneracy. By using these symmetries of
$\hat{H}$ it follows that any of its eigenstates which are
connected by a combination of these symmetry operators are
necessarily degenerate. Moreover, any eigenstates which cannot be
mapped on to each other by symmetry operators will be
non-degenerate if the symmetries of the system are
exhaustive~\cite{LL}. Other types of degeneracies can occur when
the couplings attain a particular value which indicates that at
this point the system acquires a higher symmetry as compared to
other points. This enhancement of symmetry frequently occurs in
the limits of zero or infinite coupling. An important example in
this work is when $J=0$ where $\hat{H}$ decouples into a set of
single-site Hamiltonians and acquires full permutational symmetry.
The breaking of the full permutational symmetry for finite $J$
splits the degenerate eigenstates of the $J=0$ Hamiltonian as we
will see in proceeding sections.

Particle number conservation for each component breaks up the
system Hilbert space into sectors where $\hat{N}_a\ket{\psi} =
N_a\ket{\psi}$ and $\hat{N}_b\ket{\psi} = N_b\ket{\psi}$. The
evolution of each component of an initial state in each of these
sectors can therefore be treated independently. We shall use
$\mathcal{H}_n$ to denote the sectors which possess the same total
particle number $N$ where $N_a=N-n$ and $N_b=n$ using
$n=0,1,2,\dots,N$. We denote as $\mathcal{S}_n$ the completely
symmetric subspace of the direct sum
$\mathcal{H}_n\oplus\mathcal{H}_{N-n}$, defined by
$\hat{S}\ket{\psi} = \ket{\psi}$, $\hat{R}\ket{\psi} = \ket{\psi}$
and $\hat{C}\ket{\psi} = \ket{\psi}$ for
$\ket{\psi}\in\mathcal{H}_n\oplus\mathcal{H}_{N-n}$ where
$n=0,\dots,N/2$. As we shall now outline for the specific initial
states considered here we fix $N=2M$ throughout and the only
relevant sectors are those where $n$ is restricted to be even or
$N/2$.

\subsection{Initial states and MI ground states}
Our approach to exploit the two-component BHM to engineer quantum
states is based on using two experimentally accessible initial
states. Both of these states are generated in the MI limit where
interactions are dominant, i.e. $U \gg J$ and $V \gg J$, and
precisely with two atoms localized on every site. By controlling
the spin-changing collisions a doubly occupied single-component MI
formed in some additional hyperfine state can be transformed with
near unit efficiency to the state~\cite{twoatom}
\begin{equation}
\ket{\Psi_{ab}}=\prod_{i=1}^{M}\ket{ab}_i,\label{eq:sab}
\end{equation}
with exactly the same number of $a$ and $b$ atoms in each lattice
site. Note that $\ket{\Psi_{ab}}$ is exclusively contained in the
symmetric subspace $\mathcal{S}_{M}$. By applying a further
$\pi/2$-Raman pulse to the state $\ket{\Psi_{ab}}$ an on-site
entangled MI state
\begin{equation}
\ket{\Psi_{aa+bb}} =
\prod_{i=1}^{M}\frac{1}{\sqrt{2}}(\ket{aa}_i+\ket{bb}_i),
\label{eq:sup}
\end{equation}
can be created. This state is an equal superposition of all $2^M$
possible combinations of the atom- pair states $\ket{aa}$ and
$\ket{bb}$ over $M$ sites. The states in the superposition are
contained in the symmetric subspaces $\mathcal{S}_{2 n}$ with
$n=0,1,\dots,\lfloor M/2 \rfloor$, where $\lfloor \cdot \rfloor$
denotes the integer part. Specifically, in each sector
$\mathcal{H}_{2n}$ we form the state
\begin{equation}
\ket{\psi^{2n}_{\textrm{ps}}}=\mathbbm{P}\left(\prod_{i\leq
n}\ket{aa}_i\prod_{n<j\leq M}\ket{bb}_{j}\right) \nonumber
\end{equation}
where $\mathbbm{P}(\cdot)$ stands for the normalized superposition
of all permutations of the lattice sites. Since the state
$\ket{\psi^{2n}_{\textrm{ps}}}$ possesses full permutational
symmetry it is manifestly cyclic- and mirror-symmetric. If we then
symmetrize the color as ~\footnote{Except for even $M$ where
$\kets{\Psi^{N/2}_{\textrm{ps}}}=\kets{\psi^{N/2}_{\textrm{ps}}}$.}
\begin{equation}
\ket{\Psi^{2n}_{\textrm{ps}}} \equiv
\frac{1}{\sqrt{2}}\left(\ket{\psi^{2n}_{\textrm{ps}}}+\ket{\psi^{N-2n}_{\textrm{ps}}}\right),
\label{eq:abcolor}
\end{equation}
we obtain states which are each contained in $\mathcal{S}_{2n}$,
respectively. The state $\ket{\Psi_{aa+bb}}$ can then be seen to
be a binomial superposition of the states
$\kets{\Psi^{2n}_{\textrm{ps}}}$ for all even-$n$ symmetric
subspaces as
\begin{equation}
\ket{\Psi_{aa+bb}}=\frac{1}{\sqrt{2^{M-1}}}\sum_{n=0}^{\lfloor M/2
\rfloor }\binom{M}{n}^{1/2}\ket{\Psi_{\textrm{ps}}^{2n}}.
\label{eq:aabb_state}
\end{equation}

The two initial states in Eqs.~(\ref{eq:sab}) and (\ref{eq:sup})
are exact symmetric ground states of the two-component BHM in the
limit $J=0$ for different regimes of $U/V$. This is the MI regime
where the particle-hole spectrum is gapped. In the case $U < V$
where intra-species interaction is dominant, and in the subspace
$\mathcal{S}_{N/2}$, the on-site state $\ket{ab}$ is energetically
favored making $\ket{\Psi_{ab}}$ the nondegenerate ground state.
This is easily confirmed by comparing it to all other
configurations and moreover establishes that the ground state
remains nondegenerate for non-zero $J$.

For the opposite case $U > V$ where the inter-species interaction
is dominant ``phase"-separated configurations composed of on-site
states $\ket{aa}=\ket{\circ}$ and $\ket{bb}=\ket{\bullet}$ are
energetically favored. Thus at $J=0$ the full permutational and
color symmetry results in $2 {N/2 \choose n}$ degenerate ground
states \footnote{Except for $M=N/2$ even and $n=N/4$ where the
degeneracy his given by ${N/2 \choose N/4}.$} spanned by all
configurations of the $n$ states $\ket{\circ}$ and $N/2-n$ states
$\ket{\bullet}$ distributed amongst the $M=N/2$ sites. Some of
these states can be mapped by the cyclic shift symmetry $\hat{S}$,
the mirror $\hat{R}$ or the color $\hat{C}$ transformation. We
denote as $\mathcal{G}_{2n} \subset \mathcal{S}_{2n}$ the
symmetrized (with respect to $\hat{C}$, $\hat{S}$ and $\hat{R}$)
ground state manifold of the BHM in Eq. (\ref{eq:ham}) for $J=0$
and $U > V$ with dimension denoted by $g(n,M)$.

\begin{figure}[t]
\includegraphics[width=6cm]{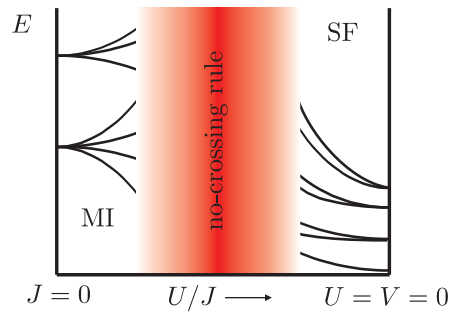}
\caption{Schematic energy-level structure for the two-component
BHM with only one varying parameter $U/J$ and fixed $U/V$. The
no-crossing rule assures that the system follows the instantaneous
eigenstates and performs no level crossings if the states have all
the same symmetry, there is only one varying parameter and the
evolution is performed sufficiently slow.}
\label{fig:nocrossingrule}
\end{figure}

On this basis we see that each of the states $\kets{\Psi^{2
n}_{\textrm{ps}}}$ composing $\ket{\Psi_{aa+bb}}$ is one of many
possible ground states in the symmetric subspace $\mathcal{G}_{2
n}$. However, as will be described in detail in
\secr{sec:small_messy}, once $J$ is non-zero this degeneracy is
partially split resulting in $\ket{\Psi_{aa+bb}}$ mapping to a
superposition of numerous low-lying excited states. This is
schematically shown in fig.~\ref{fig:nocrossingrule}. The reason
for this can be readily seen. For $0<J/U \ll 1$ nearest-neighbor
hopping will reduce the energy of configurations like
$\ket{\bullet\bullet\bullet\circ\circ\,\circ}$ where $a$ and $b$
atoms form contiguous regions, compared to configurations like
$\ket{\bullet\circ\bullet\circ\bullet\,\circ}$. Since
$\ket{\Psi_{aa+bb}}$ contains contributions from all
configurations it necessarily maps to a superposition of numerous
low-lying excited states of the two-component BHM in \eqr{eq:ham}
for each subspace $\mathcal{S}_{2 n}$ when $J$ is non-zero. This
feature significantly complicates the analysis of this state even
under adiabatic conditions.


\section{\label{sec:adiabatic}Adiabatic quantum melting}

Here we explore how adiabatic quantum melting which is an
accessible coherent process that can be applied to the system and
performs a highly non-trivial many-body unitary
transformation~\footnote{It is clear, for example, that quantum
melting of the BHM is not a single-particle process since no
product of single-particle unitaries can connect a MI state with a
non-interacting SF.} can be exploited as a means of quantum state
engineering. For both initial states $\ket{\Psi_{ab}}$ and
$\ket{\Psi_{aa+bb}}$ we take the quantum melting as ending in the
non-interacting SF limit where $U=V=0$. In this limit the
two-component BHM possesses a nondegenerate ground state
\begin{equation}
\kets{\Psi_{\rm{sf}}^{N_a,N_b}}=
\frac{(\hat{c}_0)^{N_a}(\hat{d}_0)^{N_b}}{\sqrt{N_a!N_b!}}\vac,
\label{eq:sf}
\end{equation}
for any sector $\mathcal{H}_{N_b}$ in which all the $a$ and $b$
atoms Bose-condense into their corresponding zero quasi-momentum
state.

A feasible way to analyze the properties of the final melted
states is to use the adiabatic approximation \cite{AE} in which
the system follows the instantaneous eigenstates of the
two-component BHM in \eqr{eq:ham} for each value $V/J$ and $U/J$
during the melting. The first requirement for the analysis to be
tractable is the lack of level crossings. This condition is
greatly aided by the no-crossing rule \cite{LL} which states that
if the energy levels of a quantum system are plotted as a function
of only one parameter, the curves for levels of the same symmetry
do not cross. The no-crossing rule can be followed by using the
flexibility of the optical lattice setup to ensure that the
dynamical melting only varies the ratio $V/J$ whilst keeping $U/V$
constant. Thus the analysis can proceed by following the evolution
of eigenstates in the symmetric subspace only. Also, the BHM is a
non-integrable system that shows a chaotic spectrum and thus
presents energy level repulsion \cite{IBH}. The work in \cite{IBH}
provides numerical evidence that for small systems the symmetries
we consider are exhaustive for the BHM (i.e. no crossings are
seen). We extend this assumption to arbitrary sized systems. The
energy levels may approach each other forming so-called avoided
crossings, but for infinitely slow dynamics there will be no
transitions between eigenstates. For finite times the Landau-Zener
model gives the diabatic transition probabilities
\cite{LandauZener} between two levels, and this generalizes to
Brundobler-Elser rule \cite{Brundobler} for a multistate problem.

In principle, adiabatic evolution requires that the dynamical
parameter $V/J$ changes in an infinitely long time. A crude
estimate of the time scale $t_r$ needed for adiabatic melting of
an optical lattice is given by $t_r \gg N V_{\textrm{max}}/M J^2 $
\cite{Zurek}, where $V_{\textrm{max}}/J$ stands for the initial
value in the MI regime. Importantly the adiabatic condition does
not scale with the system size but with the density $N/M$.
Numerical computations in \cite{rodriguez} indicate that the
adiabatic time scale is on the order of $t_r \sim 3
V_{\textrm{max}}/J^2$ for a finite realistically sized
one-dimensional system. These values are attainable in
experimental setups and make the adiabatic approximation a
reasonable assumption. These estimates are valid for the case when
$U=0$ when \eqr{eq:ham} decouples as two one-component BHM's.
Finding analogous time scales for the two-component system with $U
\neq 0$ is still an open problem but it is reasonable to expect
that the corresponding time scales are also attainable in
experimental setups.

\section{\label{sec:melt_ab} Melting the Product MI : $\ket{\Psi_{ab}}$}

We consider first the melting of the product MI state
$\ket{\Psi_{ab}}$. Our analysis begins by examining the outcome of
quantum melting in a small system by using an exact numerical
calculation. The results found then generalize to an arbitrarily
large system and give rise to a final state whose usefulness for
interferometry is analyzed.

\subsection{Exact numerics for a small system}
\begin{figure}[t]
\includegraphics[width=8.5cm]{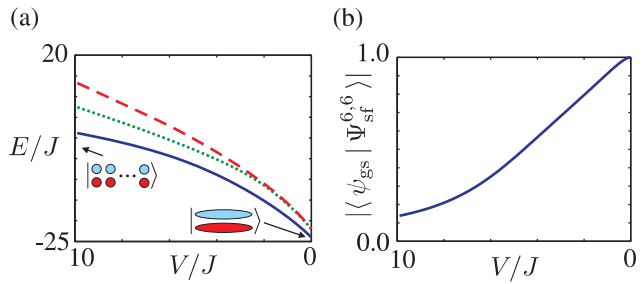}
\caption{(a) The spectrum of the two-component BHM $\hat{H}$ for
$M=6$ sites in the symmetric subspace $\mathcal{S}_6$, i.e. with
$N_a = N_b = 6$, as a function of $V/J$ with $V/U = 0.1$
throughout. To the far left (not plotted) where $V/J \gg 1$ the
state $\ket{\Psi_{ab}}$ is the non-degenerate ground state. To the
right where $V/J=0$ the non-interacting SF state $\kets{\Psi_{\rm
sf}^{6}}$ is the non-degenerate ground state. As expected these
two ground states are adiabatically connected (solid line). The
first and second excited states (dotted and dashed line,
respectively) are also shown and become degenerate in the $V=0$
limit. (b) The overlap $|\brakets{\psi_{\rm gs}}{\Psi_{\rm
sf}^{6,6}}|$ between the instantaneous ground state
$\ket{\psi_{\rm gs}}$ and the SF state $\kets{\Psi_{\rm
sf}^{6,6}}$ as a function of $V/J$.} \label{fig:small_ab_results}
\end{figure}

Here we provide some numerical evidence for the claims made in the
previous section by performing an exact numerical diagonalization
of the two-component BHM Hamiltonian $\hat{H}$ for a system with
$M=6$ sites. For the state $\ket{\Psi_{ab}}$ we need only consider
the projection of $\hat{H}$ in the subspace $\mathcal{S}_6$. We
then compute the ground state and low-lying excitations for a
sequence of linearly decreasing interaction strengths $V/J$ with a
fixed ratio $V/U = 0.1$ throughout. In
\figr{fig:small_ab_results}(a) the eigenenergies are shown
confirming that the ground state in $\mathcal{S}_6$ remains
nondegenerate during the melting, and that within the symmetric
subspace level repulsion ensures that no level crossings occur
consistent with the no-crossing rule. The final SF ground state is
then $\kets{\Psi_{\rm sf}^{6,6}}$ in which the zero quasi-momentum
mode of each component is equally occupied. Thus the quantum
melting transforms local initial correlations between pairs of
atoms into correlations between delocalized and macroscopically
occupied modes. For this reason the final state $\kets{\Psi_{\rm
sf}^{6,6}}$ is called a twin-Fock state \cite{Holland}. In
\figr{fig:small_ab_results}(b) the overlap $|\brakets{\psi_{\rm
gs}}{\Psi_{\rm sf}^{6,6}}|$ between the instantaneous ground state
$\ket{\psi_{\rm gs}}$ during the ramp and the SF twin-Fock
$\kets{\Psi_{\rm sf}^{6,6}}$ is seen to converge to unity as a
$V/J\rightarrow 0$.

\subsection{Large systems and interferometry} \label{sec:twinfock}
\begin{figure}[t]
\includegraphics[width=8.5cm]{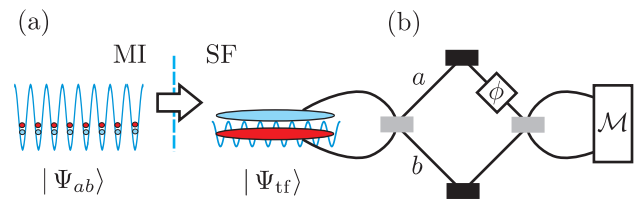}
\caption{(a) The melting of the product MI $\ket{\Psi_{ab}}$ in to
the twin-Fock SF state $\ket{\Psi_{\rm tf}}$. (b) The two internal
states $a$ and $b$ represent the arms of an interferometer with
the rotations R1, R2 and the phase-shift $\phi$ induced by
appropriate laser pulses. Finally a $\phi$ dependent observable is
measured at $\mathcal{M}$ from which $\phi$ can be deduced.}
\label{fig:mzi}
\end{figure}

We expand here the results that were investigated in
\cite{rodriguez}. By using the fact that $\ket{\Psi_{ab}}$ is a
nondegenerate MI ground state and the no-crossing rule the outcome
of adiabatic melting found for a small system can be readily
generalized to any size system as
\begin{equation}
\ket{\Psi_{ab}} \mapsto \kets{\Psi_{\rm{sf}}^{N/2,N/2}} =
\ket{\Psi_{\rm tf}}, \label{eq:tf}
\end{equation}
for arbitrary $N$. Quantum melting of the MI state
$\ket{\Psi_{ab}}$ therefore provides a direct means of creating a
$N$-particle twin-Fock state $\ket{\Psi_{\rm tf}}$, as depicted in
\figr{fig:mzi}(a).

Twin-Fock states \cite{Holland} have been proposed as useful input
states in interferometric experiments which could potentially
allow measurement sensitivities which scale as $\propto 1/N$. This
is the Heisenberg limit for interferometry and improves on the
standard limit obtained from an uncorrelated source where the
sensitivity scales as $\propto 1/\sqrt{N}$
\cite{Giovanetti04,Yurke,other}. A typical experimental scheme is
a Mach-Zenhder interferometer (MZI) that consists of an initial
beam splitter and two interferometric paths that acquire a phase
difference $\phi$ and are recombined in a final beam splitter, as
shown in \figr{fig:mzi}(b). The beam splitter operations could be
implemented by rapid resonant $\pi/2$-Raman pulses. Finally one
extracts the phase difference measuring a phase dependent
quantity.

To analyze this two-component system within the context of
interferometry we will use the Schwinger boson formalism for
constructing SU(2) operators \cite{Langular,Yurke}
\begin{eqnarray}
\hat{J}_x &=&
\frac{1}{2}\sum_i(\hat{a}_i^\dagger\hat{b}_i+\hat{b}_i^\dagger\hat{a}_i),
\nonumber \\
\hat{J}_y &=&
\frac{1}{2\ic}\sum_i(\hat{a}_i^\dagger\hat{b}_i-\hat{a}_i^\dagger\hat{b}_i),
\nonumber \\
\hat{J}_z &=& \frac{1}{2}(\hat{N}_a - \hat{N}_b), \nonumber
\end{eqnarray}
and we denote ${\bf \hat{J}} = (\hat{J_x},\hat{J_y},\hat{J_z})$.
In many cases $\hat{J}_z$ is the phase-dependent measurement made
in the interferometer. However, for the twin-Fock state
$\av{\hat{\bf J}}=0$ so a different quantity is required. One
possibility is the next order angular momentum $\hat{J}^2_z$. Note
that inefficient detection of such quantities can reduce the
sensitivities back to the standard limit \cite{Kim} and indirect
measurement schemes of $\hat{J}^2_z$ are needed
\cite{Dunningham04}.

If we use $\hat{J}^2_z$ as the phase-dependent observable the
resulting sensitivity can be calculated using error propagation
theory. This gives $\Delta \phi={\Delta \hat{J}^2_z}/{|\partial
\langle \hat{J}^2_z\rangle_\phi/\partial \phi|}$ where $\langle
\hat{J}^2_z\rangle_\phi$ and $\Delta \hat{J}^2_z
=(\av{\hat{J}^4_z}_\phi-\av{\hat{J}^2_z}_\phi)^{1/2}$ are the
average and the spread respectively. To calculate
$\av{\cdot}_\phi$ one can either rotate the state or the operators
\cite{Yurke}. Since the action of the MZI is equivalent to the
unitary operation $\exp(-i \phi \hat{J}_y)$, applying it to the
operators gives
\begin{eqnarray}
\av{\hat{J}^2_z}_\phi &=& \av{\hat{J}^2_z} \cos^2(\phi)
+\av{\hat{J}^2_x} \sin^2(\phi) \nonumber \\
&& -\, \sin(\phi) \cos(\phi) \av{\{\hat{J}_z,\hat{J}_x\}},
\nonumber
\end{eqnarray}
and
\begin{eqnarray}
\av{\hat{J}^4_z}_\phi &=& \av{\hat{J}^4_z} \cos^4(\phi) +
\av{\hat{J}^4_x} \sin^4(\phi) \nonumber \\
&& +\, \sin^2(\phi)\cos^2(\phi)\av{\{\hat{J}_x \hat{J}_z,\hat{J}_z
\hat{J}_x\}
\nonumber \\
&& +\, \{\hat{J}_x^2,\hat{J}_z^2\} + \{\hat{J}_x,\hat{J}_z \hat{J}_x \hat{J}_z\}} \nonumber \\
&& -\,\cos^3(\phi)\sin(\phi)\av{\{\hat{J}_x,\hat{J}_z^3\} + \{\hat{J}_z,\hat{J}_z\hat{J}_x\hat{J}_z\}} \nonumber \\
&& -\, \sin^3(\phi)\cos(\phi)\av{\{\hat{J}_z,\hat{J}_x^3\} +
\{\hat{J}_x,\hat{J}_x\hat{J}_z\hat{J}_x\}}, \nonumber
\end{eqnarray}
where $\{\hat{A},\hat{B}\}=\hat{A}\hat{B} + \hat{B}\hat{A}$ is the
anticommutator. For states which are zero eigenvectors of
$\hat{J}_z$, such as twin-Fock states, this yields
\begin{equation}
\Delta \phi^2=\frac{\sin^2\phi(\langle \hat{J}_x^4 \rangle-\langle
\hat{J}_x^2 \rangle^2)+ \cos^2 \phi \langle \hat{J}_x \hat{J}_z^2
\hat{J}_x \rangle}{4 \cos^2 \phi \langle \hat{J}_x^2 \rangle^2}.
\nonumber
\end{equation}
The minimum value of this quantity occurs at $\phi=0$ and the
sensitivity reduces to $\Delta \phi({0})=\half\langle \hat{J}_x^2
\rangle^{-1/2}$ after using the equality $\langle \hat{J}_x
\hat{J}_z^2 \hat{J}_x \rangle =\langle \hat{J}_x^2 \rangle$ which
holds for such states.

The sensitivity can be expressed entirely in terms of the
one-particle density matrices ${\rho}^{a}_{ij} = \langle
\hat{a}^{\dagger}_i\hat{a}_j \rangle$ and ${\rho}^{b}_{ij} =
\langle \hat{b}^{\dagger}_i\hat{b}_j \rangle$ using
\begin{equation}
\langle \hat{J}_x^2 \rangle = \frac{1}{4}\left[\sum_i(\rho^a_{ii}
+ \rho^b_{ii}) + \sum_{i,j}(\rho^a_{ij}\rho^b_{ji} +
\textrm{h.c.})\right]. \label{eq:jx2}
\end{equation}
For the initial (uncorrelated) MI state $\ket{\Psi_{ab}}$, which
possesses no off-diagonal correlations
${\rho}^a_{ij}={\rho}^b_{ij}=\delta_{ij}$, this yields a
sensitivity at the standard limit $\Delta \phi=1/{\sqrt{2 N}}$.
For the final SF $\ket{\Psi_{\textrm{sf}}}$, which possesses
long-range correlations $\rho^a_{ij}=\rho^b_{ij}=1$, we see that
asymptotically the Heisenberg limit $\Delta \phi=1/{\sqrt{N^2/2 +
N}}$ is recovered. Thus the scaling $\Delta \phi \propto
N^{-\alpha}$ changes from $\alpha=1/2$ to $\alpha=1$ during the
melting. An inspection of Eq.~(\ref{eq:jx2}) shows that one needs
non-zero values of the long-range correlations $\rho_{ij}$ in
order to have better scaling than $N$ given by the first term in
Eq.~(\ref{eq:jx2}). We showed in \cite{rodriguez} that one can
obtain sensitivities approaching the HL scaling even in the
presence of depletion if $U=0$. Recent results in \cite{BdG}
indicate that our conclusions about the effect of depletion are
also valid for $U>0$. Moreover, we also showed in \cite{rodriguez}
that one can obtain HL sensitivities ($\alpha \sim 1$) if
particles are lost during the melting process.


\section{Melting the Entangled MI : $\ket{\Psi_{aa+bb}}$}
\label{sec:messy}

In this section we analyze the melting of the entangled MI state
$\ket{\Psi_{aa+bb}}$. In previous work \cite{conf} we demonstrated
that this state may melt into a maximally entangled state if the
lattice is infinitely connected. In a typical experimental set-up
with only nearest neighbor hopping, the initial permutational
symmetry for $J=0$ is broken for finite $J$ and the final SF state
is highly excited. We begin by again considering the outcome of
quantum melting in a small system. Finally after applying a number
of approximations we compute the momentum correlations of the
final melted state for a 1D lattice.

\subsection{Small system}
\label{sec:small_messy}

For $M=6$ the entangled MI state $\ket{\Psi_{aa+bb}}$ is an equal
superposition of all 64 possible combinations of the atom-pair
states $\ket{\circ}$ and $\ket{\bullet}$ over $M$ sites. Following
\eqr{eq:aabb_state} it can be written as
\begin{eqnarray}
  \ket{\Psi_{aa+bb}} &=& \frac{1}{8}\left(\sqrt{2}\ket{\Psi_{\rm ps}^0} + \sqrt{12}\ket{\Psi_{\rm ps}^2} + \sqrt{30}\ket{\Psi_{\rm ps}^4}\right. \nonumber \\
  && +\, \left.\sqrt{20}\ket{\Psi_{\rm ps}^6}\right), \nonumber
\end{eqnarray}
where the phase-separated states in each subspace
$\mathcal{S}_{2n}$ are
\begin{eqnarray}
  \ket{\Psi_{\rm ps}^0} &=& \mathbbm{S}\left(\ket{\bullet\bullet\bullet\bullet\bullet\,\bullet}\right), \nonumber \\
  \ket{\Psi_{\rm ps}^2} &=& \mathbbm{S}\left(\ket{\bullet\bullet\bullet\bullet\bullet\,\circ}\right), \nonumber \\
  \ket{\Psi_{\rm ps}^4} &=& \frac{1}{\sqrt{30}}\left[\sqrt{6}\,\mathbbm{S}\left(\ket{\bullet\bullet\circ\bullet\bullet\,\circ}\right) + \sqrt{12}\,\mathbbm{S}\left(\ket{\bullet\bullet\bullet\circ\bullet\,\circ}\right)\right. \nonumber \\
  && +\, \left.\sqrt{12}\,\mathbbm{S}\left(\ket{\bullet\bullet\bullet\bullet\circ\,\circ}\right)\right], \nonumber \\
  \ket{\Psi_{\rm ps}^6} &=& \frac{1}{\sqrt{20}}\left[\sqrt{2}\,\mathbbm{S}\left(\ket{\bullet\circ\bullet\circ\bullet\,\circ}\right) +
  \sqrt{12}\,\mathbbm{S}\left(\ket{\bullet\bullet\circ\bullet\circ\,\circ}\right)\right.
  \nonumber\\
  && +
  \,\left.\sqrt{6}\,\mathbbm{S}\left(\ket{\bullet\bullet\bullet\circ\circ\,\circ}\right)\right].\label{state_small}
\end{eqnarray}
Here we have used $\mathbbm{S}(\cdot)$ to denote the normalized
equal superposition of all equivalent configurations under the
symmetry operations $\hat{S}, \hat{R}$ and $\hat{C}$. Each phase
separated state $\ket{\Psi_{\rm ps}^{2n}}$ is a member of
$\mathcal{G}_{2 n}$ and is a superposition of all the $g$
inequivalent symmetric states within it (note that
$g(0,6)=g(1,6)=1$, while $g(2,6)=g(3,6)=3$). The weights of this
superposition Eq.~(\ref{state_small}) are given by the
degeneracies due to symmetrization of these inequivalent states,
which we denote by $s(l)$ where $l=1,.,g$.

\begin{figure}[t]
\includegraphics[width=8.5cm]{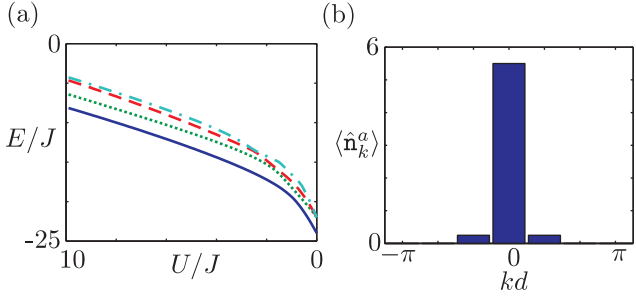}
\caption{(a) The low-energy spectrum of the two-component BHM
$\hat{H}$ for $M=6$ sites in the symmetric subspace
$\mathcal{S}_4$ as a function of $U/J$ with $V/U=0.1$ throughout.
The initial state $\kets{\Psi_{aa + bb}}$ projected in this
subspace is $\ket{\Psi_{\rm ps}^4}$. The adiabatic evolution of
this state involves the ground state and two lowest-lying excited
states. In the SF limit with $U=0$ the first-excited level is
triply degenerate. Consistent with the no-crossing rule we see
that energy levels within $\mathcal{S}_4$ repel each other and
only avoided crossings are seen. (b) The momentum distribution
$\av{\hat{{\tt n}}^a_k}$ for one component (it does not matter
which due to color symmetry) of the final adiabatically melted
state $\ket{\Psi_{\rm f}}$ derived from $\kets{\Psi_{aa + bb}}$
with $M=6$.} \label{fig:small_messy}
\end{figure}

We now proceed to melt each of these phase-separated contributions
separately. For the subspaces $\mathcal{S}_0$ and $\mathcal{S}_2$
the corresponding states $\ket{\Psi^0_{\rm ps}}$ and
$\ket{\Psi^2_{\rm ps}}$ are non-degenerate MI ground states.
Analogous to the melting of $\ket{\Psi_{ab}}$ in the subspace
$\mathcal{S}_6$ exact numerical diagonalization of the
two-component BHM with $U>V$ confirms that there are no level
crossings and so these two states are adiabatically connected to
their corresponding symmetric SF ground states
\begin{eqnarray}
  \mathbbm{S}\left(\ket{\bullet\bullet\bullet\bullet\bullet\,\bullet}\right) &\mapsto& \frac{1}{\sqrt{2}}\left(\kets{\Psi_{\rm
  sf}^{12,0}} + \kets{\Psi_{\rm
  sf}^{0,12}}\right), \nonumber \\
  \mathbbm{S}\left(\ket{\bullet\bullet\bullet\bullet\bullet\,\circ}\right) &\mapsto& \frac{1}{\sqrt{2}}\left(\kets{\Psi_{\rm
  sf}^{10,2}} + \kets{\Psi_{\rm
  sf}^{2,10}}\right). \nonumber
\end{eqnarray}
In this notation $\mapsto$ we omit the dynamical phase $e^{i
\alpha}$ acquired by the state during adiabatic melting
\cite{conf}. In contrast subspaces $\mathcal{G}_{4}$ and
$\mathcal{G}_{6}$ are three-fold degenerate and the states
$\ket{\Psi^4_{\rm ps}}$ and $\ket{\Psi^6_{\rm ps}}$ are both
composed of 3 inequivalent MI configurations. This situation
becomes more typical with increasing system size. In the $J=0$
limit these configurations are degenerate, however once $J$ is
non-zero they split in energy. In \figr{fig:small_messy}(a) the
energies for the ground state and three lowest lying excitations
in $\mathcal{S}_4$ are shown for a linear ramping of $U/J$ with
$U/V=0.1$. This, together with a similar result for
$\mathcal{S}_6$ (not shown), confirm that no level crossings
occur. For this reason we can map the MI configurations via their
ordering in energy with their corresponding ground state and
excitations in the SF limit. The lowest energy configurations
adiabatically connected to the symmetric SF ground state are
\begin{eqnarray}
  \mathbbm{S}\left(\ket{\bullet\bullet\bullet\bullet\circ\,\circ}\right) &\mapsto& \frac{1}{\sqrt{2}}\left(\kets{\Psi_{\rm
  sf}^{8,4}} + \kets{\Psi_{\rm
  sf}^{4,8}}\right), \nonumber \\
  \mathbbm{S}\left(\ket{\bullet\bullet\bullet\circ\circ\,\circ}\right) &\mapsto& \kets{\Psi_{\rm
  sf}^{6,6}}, \nonumber
\end{eqnarray}
respectively, while the two remaining configurations map to the
lowest and second-lowest excitations in the SF limit. Quite
generally there are many degenerate SF excitations in the $U=V=0$
limit. For the subspace $\mathcal{S}_6$ the first two lowest-lying
excitations are degenerate, whereas the first three are for
$\mathcal{S}_4$. To determine the appropriate state to which each
configuration maps to we apply first-order perturbation theory
(see \appr{app:sfpert}) in the inter- and intra-species
interaction which splits these states and identifies the correct
excitations. This then gives for $\mathcal{S}_6$
\begin{eqnarray}
  \mathbbm{S}\left(\ket{\bullet\bullet\circ\bullet\circ\,\circ}\right) &\mapsto& \kets{\Psi_{{\rm
  ex},1}^{6}}, \nonumber \\
  \mathbbm{S}\left(\ket{\bullet\circ\bullet\circ\bullet\,\circ}\right) &\mapsto& \kets{\Psi_{{\rm
  ex},2}^{6}}, \nonumber
\end{eqnarray}
and for $\mathcal{S}_4$ we have
\begin{eqnarray}
  \mathbbm{S}\left(\ket{\bullet\bullet\bullet\circ\bullet\,\circ}\right) &\mapsto& \kets{\Psi_{{\rm
  ex},1}^{4}}, \nonumber \\
  \mathbbm{S}\left(\ket{\bullet\bullet\circ\bullet\bullet\,\circ}\right) &\mapsto& \kets{\Psi_{{\rm
  ex},2}^{4}}, \nonumber
\end{eqnarray}
where $\kets{\Psi_{{\rm ex},1}^n}$ and $\kets{\Psi_{{\rm
ex},2}^n}$ denote the lowest and second-lowest SF excitations in
$\mathcal{S}_n$ (see \appr{app:sfpert}). Note that this
calculation of the lowest excitations works for arbitrary $M$,
although it becomes cumbersome for larger system size as we would
need a very large number of SF excitations.

\begin{figure}[t]
\includegraphics[width=8cm]{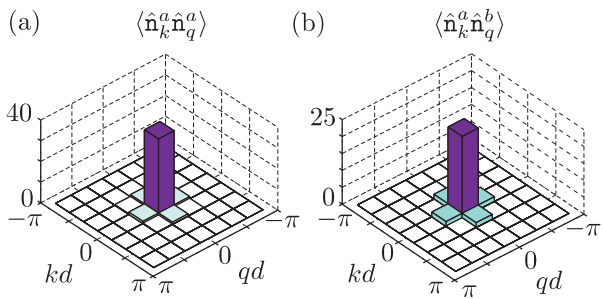}
\caption{(a) Momentum space particle number correlations
$\av{\hat{{\tt n}}^a_k \hat{{\tt n}}^a_q}$ of the final melted
state $\ket{\Psi_{\rm f}}$ for one component (it does not matter
which due to color symmetry) for $M=6$. (b) The inter-species
momentum space particle number correlations $\av{\hat{{\tt n}}^a_k
\hat{{\tt n}}^b_q}$ for $\ket{\Psi_{\rm f}}$.}
\label{fig:mom_aa_ab}
\end{figure}

Combining these results we can write the overall final SF state
which is adiabatically connected to $\ket{\Psi_{aa+bb}} \mapsto
\ket{\Psi_{\rm f}}$ with $M=6$ as
\begin{widetext}
\begin{eqnarray}
  \ket{\Psi_{\rm f}} &=&\frac{1}{8}\left[e^{i\alpha_{0}}\left(\kets{\Psi_{\rm
  sf}^{12,0}} + \kets{\Psi_{\rm
  sf}^{0,12}}\right) + \sqrt{6}e^{i\alpha_{2}}\left(\kets{\Psi_{\rm
  sf}^{10,2}} + \kets{\Psi_{\rm
  sf}^{2,10}}\right) + \sqrt{6}e^{i\alpha_{4}}\left(\kets{\Psi_{\rm
  sf}^{8,4}} + \kets{\Psi_{\rm
  sf}^{4,8}}\right) + \sqrt{12}
  e^{i\alpha_{4,1}}\kets{\Psi_{{\rm
  ex},1}^{4}} \right. \nonumber \\
  && \,+ \left. + \sqrt{6} e^{i\alpha_{4,2}}\kets{\Psi_{{\rm
  ex},2}^{4}}+\sqrt{6} e^{i\alpha_{6}}\kets{\Psi_{\rm
  sf}^{6,6}} + \sqrt{12} e^{i\alpha_{6,1}}\kets{\Psi_{{\rm
  ex},1}^{6}} + \sqrt{2} e^{i\alpha_{6,2}}\kets{\Psi_{{\rm
  ex},2}^{6}} \right]. \nonumber
\end{eqnarray}
\end{widetext}
where we have now written explicitly the dynamical phases acquired
by each of the states in the superposition. An observable of
experimental significance is the momentum distribution for either
component. In fact the momentum distribution $\av{\hat{{\tt
n}}_{\pm k}^a} = \av{\hat{{\tt n}}_{\pm k}^b}$ is not dependent on
the precise form of the excited states $\ket{\Psi_{\rm ex}}$ since
all states in the same degenerate manifold have the same
expectation value ($\av{\hat{{\tt n}}_{\pm \frac{\pi}{3d}}^a} =
\half$ in this case). The resulting momentum distribution is shown
in \figr{fig:small_messy}(b), and gives the fraction of the
population outside the zero-momentum mode (i.e. depletion of the
Bose-condensate) for either component as $D = 1/12 \approx 8.3\%$.
Both the intra-species, $\av{\hat{{\tt n}}_{\pm k}^a\hat{{\tt
n}}_{\pm q}^a}$, and the inter-species, $\av{\hat{{\tt n}}_{\pm
k}^a\hat{{\tt n}}_{\pm q}^b}$, momentum space particle number
correlations do depend on the precise form of the excited states
$\ket{\Psi_{\rm ex}}$. Neither the momentum distributions nor the
momentum space particle number correlations depend on the relative
dynamical phases. In \figr{fig:mom_aa_ab}(a)(b) these correlations
are plotted using the excited states determined from first-order
perturbation theory with $U/J=0.01$ and $V/U = 0.1$.

\subsection{Large systems}
\label{sec:nn_latt}

The final SF state obtained via adiabatic melting the entangled MI
state $\ket{\Psi_{aa+bb}}$ is given by
\begin{equation}\label{eq:psiaabbm}
\ket{\Psi_{\textrm{f}}}=\frac{1}{\sqrt{2^{M-1}}}\sum_{n=0}^{\lfloor
M/2 \rfloor}\binom{M}{n}^{1/2}\ket{{\Psi}_{\textrm{f}}^{2n}},
\end{equation}
where $\ket{{\Psi}_{\textrm{ps}}^{2n}} \mapsto
\ket{{\Psi}_{\textrm{f}}^{2n}}$ if we assume that $U/V$ is fixed
during melting and that the symmetries considered are exhaustive
and thus no level crossings occur. The numerical calculation for
$M=6$ in section \ref{sec:small_messy} hints that it is very
complicated to obtain the final state for larger $M$ which are
inaccessible with exact numerics. Resolving the exact splitting of
the states in each subspace $\mathcal{S}_{2n}$ in the MI limit and
in the SF limit is both complex and impractical for an arbitrarily
large $M$. Instead, we introduce here some approximations that
allow us to calculate the momentum distributions and momentum
correlations of the final states.

In the MI regime for $J=0$ and $U>V$ the color-symmetric ground
state manifold is $\binom{M}{n}$ degenerate. For finite $J$, the
degeneracy is broken and $g(n,M)$ states (obtained after
symmetrization in $\hat{S}, \hat{R}$) are not coupled by the
evolution. As shown below for large $M$ we can approximate the
initial MI state $\ket{{\Psi}_{\textrm{ps}}^{2n}}$ as an equal
superposition of the $g(n,M)$ states. In this way, using the
adiabatic approximation we can map these $g$ states into the
lowest $g$ states in the SF regime.  We calculate the lowest
eigenstates of the BHM in the SF regime $U=V=0$ and their
degeneracies $f(l)$ such that $\sum_l f(l)=g$. Even without
calculating the exact coefficients of the final $g$ states in the
SF eigenbasis, we can compute the momentum distribution and the
momentum space particle number correlations of the final melted
state to good approximation.

\subsubsection{MI limit}

If $U/V>1$ and $J=0$ the phase-separated state
$\ket{\Psi_{\textrm{ps}}^{2n}}$ belongs to the ground state
manifold $\mathcal{G}_{2n} \subset \mathcal{S}_{2n}$. It is
constructed as an equal superposition of the $\binom{N/2}{n}$
atom-pair states $\ket{\circ}$ and $\ket{\bullet}$ distributed
over $M=N/2$ sites. These atom-pair states belong to the ground
state manifold. However, only $g(n,M) \leq \binom{N/2}{n}$ of them
define the symmetrized ground-state manifold $\mathcal{G}_{2n}$
and are not coupled by evolution for finite $J$. We showed in
section \ref{sec:small_messy} that e.g. $g(3,6)=3$ and that for
$M=6$, the state $\ket{{\Psi}_{\textrm{ps}}^{6}}$ is a
superposition of $3$ states that evolve into $3$ different final
SF states. In general, we can calculate the dimension of the
subspace $\mathcal{G}_{2 n}$, $g(n,M)$, by noticing that it is
isomorphic to the number of color symmetric configurations in a
bracelet (c.f. Fig.\ref{fig:sym}) with $n$ beads of one color and
$M-n$ beads of the other color. Such mathematical structures have
been extensively studied \cite{seriesequal,seriesunequal} and
there are analytical formulas available for their properties. For
even $M$, the number of complementable bracelets with $M$ beads
and half of them of each color is given by \cite{seriesequal}
\begin{eqnarray}
& g(M/2,M)=\frac{1}{2}
[F(\frac{M}{2})+2^{\frac{M}{2}-2}+\binom{\frac{M}{2}-1}{\frac{M-2}{4}}]
\hspace{2mm} \frac{M}{2} \hspace{1mm} \textrm{odd} \nonumber \\
& g(M/2,M)=\frac{1}{2}
[F(\frac{M}{2})+2^{\frac{M}{2}-2}+\binom{\frac{M}{2}}{\frac{M}{4}}]
\hspace{2mm} \frac{M}{2} \hspace{1mm} \textrm{even},\nonumber
\end{eqnarray}
where $F(n)=\frac{1}{2n}\sum_{d|n}\Phi (n/d)\binom{2
d-1}{d-1}+\Phi (2n/d) 2^{d-1} $ and $\Phi(x)$ is Euler's totient
function. For bracelets with odd total number of beads $M$ or even
$M$ and $n \neq M/2$, the number of complementable bracelets is
given by \cite{seriesunequal}
\begin{eqnarray}
& g(n,M)=\frac{1}{2} [\textrm{Fold}({G,0,\{d: d|M \cap d|n
\}})+H(n,M)]. \nonumber
\end{eqnarray}
Here, $\textrm{Fold}(G(\cdot,\cdot),x,\{a_1,\dots,a_k \})$ gives
the last term of the list $\{x,G(x,a_1),G(G(x,a_1),a_2),\dots \}$
generated by the cumulative application of the function
\begin{equation}
G(x,y)=x+\frac{\phi(y)}{M}\binom{\frac{M}{y}}{\frac{n}{y}}
\nonumber
\end{equation}
to the elements of the input list $\{a_1,\dots,a_k\}$ and
$$
H(n,M) = \left\{ \begin{array}{rl}
 \binom{\frac{M-2}{2}}{\frac{n-1}{2}} &\mbox{ if $M$ even, $n$ odd} \\
 \binom{\lfloor n/ 2\rfloor }{\lfloor M/2\rfloor } &\mbox{ otherwise.} \\
       \end{array} \right.
$$
 Note
that we recover $g(2,6)=g(3,6)=3$ obtained for the states
$\ket{\Psi_{\textrm{ps}}^{4}}$ and $\ket{\Psi_{\textrm{ps}}^{6}}$
given in Sec.\ref{sec:small_messy}.

\subsubsection{Superposition states}
\begin{figure}[t]
\includegraphics[width=8.5cm]{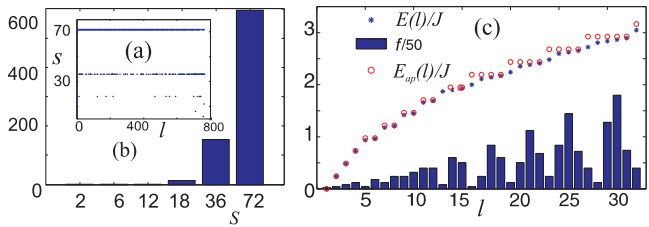}
\caption{(a) Number of MI ground states ($d_{MI}$) that belong to
each of the $g(9,18)=765$ independent states for finite $J$ for a
lattice with $N/2=N_a=N_b=18$ and $U>V$. (b) Number of independent
states with a given degeneracy $d_{MI}$. The total number of
independent states is equal to $g(9,18)=765$. (c) First energy
levels ($\ast$) in the SF regime ($U=V=0$) and their degeneracy
$f$ (bar plot) for $M=18$. The points ($\circ$) show the quadratic
approximation to the dispersion relation.  We plot only the first
states such that $\sum_l f(l)=g$ for $g=765$.}\label{fig:deg}
\end{figure}

We now need to write our initial state
${\ket{\Psi_{\textrm{ps}}^{2n}}}$ as a superposition of the
$g(n,M)$ states that evolve independently for finite $J$. Using
the no-crossing rule we can map the initial superposition
$\ket{\Psi_{\textrm{ps}}^{2n}}$ at $J=0$ into a final
superposition $\ket{{\Psi}^{2n}_{\textrm{f}}}$. The coefficients
of the superposition are given by the the number of bracelet
configurations $s(l)$ that can be obtained by mirror inversion
$\hat{R}$ or cyclic permutation $\hat{S}$ of any of the
independent states $l=1,..,g(n,M)$ in the subspace
$\mathcal{G}_{2n}$. We show $s(l)$ for $M=18$ in \figr{fig:deg}
(a). Note that $\sum_l^{g} s(l)=\binom{M}{n}$. Numerical results
in \figr{fig:deg} (b) show that most ($77\%$) of the $g$
inequivalent configurations have the same degeneracy $s$ and that
this behavior increases with size $M$ (for $M=22$ already $92\%$
of the $g$ bracelets present the same degeneracy) . We can thus
assume that the number of states isomorphic to each independent
configuration is constant $s=\binom{M}{n}/g$. Within this
approximation, $\ket{{\Psi}^{2n}_{\textrm{f}}}$ is thus an equal
superposition of the first $g$ eigenstates of the SF Hamiltonian.

\subsubsection{SF limit}
In Appendix B we show how to calculate the eigenstates $\ket{\{
e_l \}}$ of the Hamiltonian (\ref{eq:ham}) in the SF regime
 where $U=V=0$. The lowest eigenstates in ascending order in energy for
large $M$ can be calculated efficiently assuming a quadratic
approximation $E_{ap}(l)$ to the non-interacting single-particle
energy $E(l)$. Due to the color symmetry, the SF eigenstates
$\ket{e_l}$ are degenerate. The degeneracies of the states $f(l)$
can be calculated numerically. We show in \figr{fig:deg} (c) the
eigenenergies of the first SF states and their degeneracies for
$M=18$.

To obtain the momentum distribution of
$\kets{\Psi^{2n}_{\textrm{f}}}$ we do not need to compute the
exact distribution of the final states
$\ket{\Psi_{\textrm{ex},m}}$ in the $\{\ket{e_l}\}$ basis (see the
Appendix \ref{app:sfpert}) because all states in the same
degenerate manifold have the same expectation value $\av{\hat{{\tt
n}}_q^a}=\av{\hat{{\tt n}}_q^b}$. We just need to calculate the
momentum distribution of the final eigenstates $\{\ket{e_l}\}$ and
their degeneracies $f(l)$ and sum their contributions up to the
highest excited state such that $\sum_l f(l)\approx g$.

On the other hand, we saw in Sec.~\ref{sec:small_messy} that we
need the exact form of the final states
$\ket{\Psi_{\textrm{ex},m}}$ in order to obtain the
momentum-number correlations. Note, however, that
$\sum_{m=1}^{f}\bra{e_{m}} \hat{\tt n}_k \hat{\tt n}_q
\ket{e_m}=\sum_{m=1}^{f}\bra{\Psi_{\textrm{ex},m}} \hat{\tt n}_k
\hat{\tt n}_q \ket{\Psi_{\textrm{ex},m}}$ where we sum over a
complete degenerate manifold. Thus, if we make the further
approximation that the state $\ket{{\Psi}^{2n}_{\textrm{f}}}$
fully populates the highest excited manifold we can calculate its
momentum space particle number correlations without knowing the
exact states $\ket{\Psi_{\textrm{ex},m}}$. This is justified for
large $g$ because the exact form of the highest excited states is
concealed by the lower manifolds that are fully populated.

\begin{figure}[t]
\includegraphics[width=8.5cm]{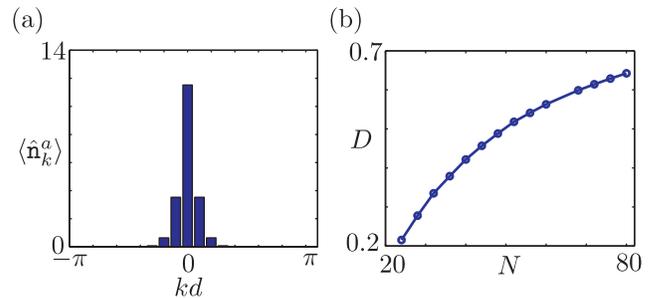}
\caption{ (a) The momentum distribution of the final SF state
$\ket{\Psi_{\textrm{f}}}$ obtained via adiabatically melting the
MI state $\ket{\Psi_{\textrm{aa+bb}}}$ with $N_a=N_b=M=20=N/2$ and
$U/V>1$. (b) The depletion $D = 1 - \av{{\tt n}^a_0}/N_a$ of the
melted state $\kets{\Psi^{N/2}_{\textrm{f}}}$ for $U/V>1$ as a
function of the number of particles $N$. The solid line shows the
best logarithmic fit $-0.0748 (\ln N/2)^2+0.8201 \ln N/2-1.3633$
to the calculated values (shown as $\circ$).} \label{fig:MOM}
\end{figure}

We show the momentum distribution of the melted state
$\ket{\Psi_{\textrm{f}}}$ in \figr{fig:MOM} (a). The final state
is highly depleted. A plot of the depletion $D=1 - \av{{\tt
n}_0}/N$ of the term $\kets{\Psi^{N/2}_{\textrm{f}}}$ as a
function of $N$ is shown in Fig. \ref{fig:MOM} (b). It indicates
the behavior of the depletion of the full state
$\ket{\Psi_{\textrm{f}}}$ because this term has the highest
binomial coefficient in the sum (\ref{eq:psiaabbm}) and a very
large $g$. One can see that the depletion increases very rapidly
with the number of particles.

\begin{figure}[t]
\includegraphics[width=8cm]{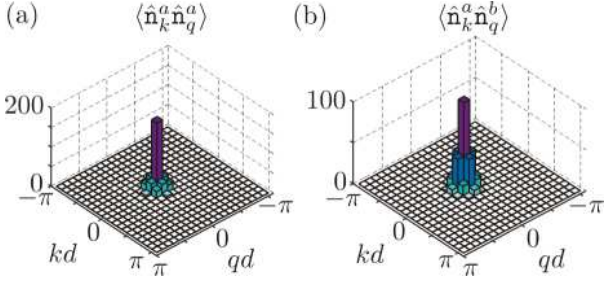}
\caption{ (a) The momentum correlations $\av{\hat{{\tt n}}^a_k
\hat{{\tt n}}^a_q}$ for one mode for the state $\ket{\Psi_{\rm
{f}}}$ obtained via the adiabatic melting of $\ket{\Psi_{aa+bb}}$
with $N=40$ and $U/V>1$. (b) The inter-species momentum
correlations $\av{\hat{{\tt n}}^a_k \hat{{\tt n}}^b_q}$ of
$\ket{\Psi_{\rm {f}}}$ with $N=40$. Note that a twin-Fock state
(\ref{eq:tf}) would only show a central peak in both
distributions.}\label{fig:MOMC}
\end{figure}

The momentum-space particle number correlations of the melted
state $\ket{\Psi_{\textrm{f}}}$ are shown in figure
\ref{fig:MOMC}. We observe that the population outside the zero
momentum mode is higher for the inter-species momentum
correlations. Our calculations clearly indicate that melting two
component Mott insulators leads to quantum states with a
complicated structure and long range correlations. A number of
approximations were necessary to arrive at these results. We
believe that an exact treatment is unlikely to reveal a simple
structure which allows an intuitive interpretation of the melting
process.

\section{Summary}
We have analyzed the quantum melting of correlated two-component
MI states into the SF regime using the adiabatic approximation.
Based on the experimental fact that the one component MI-SF
transition can be performed near adiabatically, we have applied it
to the two-mode case. Using the no-crossing rule we have shown
that quantum melting of a two-component MI in an optical lattice
provides a viable route for engineering twin Fock states useful in
interferometric experiments. The twin Fock states present only
macroscopic occupation of the final SF ground state although in
general this is not the case. We have shown that due to the
breaking of the translational symmetry when there is nearest
neighbor hopping the final SF connected to a ground initial state
can be highly depleted even in the limit of adiabatic behavior.

Using a set of well justified approximations, we have calculated
the momentum correlations and momentum space particle number
correlations for the final states obtained via melting on-site
entangled states created in the MI regime. Our results clearly
indicate that the melted SF states present a complicated structure
and long range correlations. Further work is needed to ascertain
if these correlations can be exploited for interferometry or other
tasks.

 \section*{Acknowledgements}
 M. R. thanks K. Surmacz, J.J. Garc\'ia-Ripoll and R. A. Molina for fruitful discussions and the Spanish MCyT
under programme Juan de la Cierva for support. This research was
supported by the EPSRC projects EP/E041612/1 and EP/C51933/1, and
by the National Science Foundation under Grant No. NSF
PHY05-51164. M. R. and D. J acknowledge support from the EU
through the STREP project OLAQUI.

\appendix
\section{Splitting degenerate SF excitations}
\label{app:sfpert}


In the non-interacting SF limit where $U=V=0$ the excited states
are made up by expelling pairs of atoms of opposite momentum from
the ground state Eq.~(\ref{eq:sf}). Of particular relevance to the
exact calculation of the melting of $\ket{\Psi_{aa+bb}}$ in a
small system are excitations in the subspaces $\mathcal{S}_{M}$,
where $|N_a - N_b| = 0$, and $\mathcal{S}_{M-2}$, where $|N_a -
N_b| = 4$ and $M=N/2$. Here we shall demonstrate that the
lowest-lying excitations in $\mathcal{S}_{M}$ and
$\mathcal{S}_{M-2}$ are split by the inter- and intra-species
interaction to first-order in $V/J$ and $U/J$.

For the subspace $\mathcal{S}_M$ the two lowest-lying excitations
are degenerate. These form an excitation subspace which is spanned
by states where a single pair of atoms, of either component, is
expelled from the condensate to equal and opposite momenta with
the lowest possible magnitude as
\begin{eqnarray}
  \ket{e_1} &=& \frac{1}{\sqrt{2}}\left(c_+^{\dagger}d_-^{\dagger}+
  c_-^{\dagger}d_+^{\dagger}\right)\kets{\Psi_{\rm
  sf}^{M-1,M-1}},\nonumber
  \\
  \kets{e_2} &=& \frac{1}{\sqrt{2}}\left(c_+^{\dagger}c_-^{\dagger}\kets{\Psi_{\rm sf}^{M-2,M}}
  + d_-^{\dagger}d_+^{\dagger}\kets{\Psi_{\rm sf}^{M,M-2}}\right),
 \label{eq:exc}
\end{eqnarray}
where $\pm$ is used to denote the smallest quasi-momenta
$\pm\delta k$. The inter- and intra-species interaction matrix
elements in this excitation subspace are
\begin{displaymath}
\left[\begin{array}{cc}
V(M+1-\frac{2}{M}) + U(M+\frac{1}{M}) & 2U\sqrt{1-\frac{1}{M}} \\
2U\sqrt{1-\frac{1}{M}} & V(M+1-\frac{3}{M}) + U M
\end{array}\right].
\end{displaymath}
Upon diagonalizing this $2 \times 2$ matrix the resulting gap and
non-degenerate excitations $\ket{\Psi_{{\rm ex},1}^M}$ and
$\ket{\Psi_{{\rm ex},2}^M}$ (which will in general be some
superposition of the states $\ket{e_1}$ and $\ket{e_2}$) are
obtained in first-order perturbation theory.

In a similar way for subspace $\mathcal{S}_{M-2}$ the three
lowest-lying excitations are degenerate. This excitation subspace
is again spanned by states where a single pair of atoms is
expelled from the condensate to equal and opposite momenta with
the lowest possible magnitude as
\begin{eqnarray}
  \ket{e_1} &=& \frac{1}{\sqrt{2}}\left(c_+^{\dagger}c_-^{\dagger}\kets{\Psi_{\rm sf}^{M-4,M+2}}+
  d_-^{\dagger}d_+^{\dagger}\kets{\Psi_{\rm sf}^{M+2,M-4}}\right), \nonumber \\
  \ket{e_2} &=& \frac{1}{2}\left(c_+^{\dagger}d_-^{\dagger} + c_-^{\dagger}d_+^{\dagger}\right)\left(\kets{\Psi_{\rm sf}^{M+1,M-3}} + \kets{\Psi_{\rm sf}^{M-3,M+1}}\right), \nonumber \\
  \ket{e_3} &=& \frac{1}{\sqrt{2}}\left(c_+^{\dagger}c_-^{\dagger}\kets{\Psi_{\rm sf}^{M,M-2}}+
  d_-^{\dagger}d_+^{\dagger}\kets{\Psi_{\rm sf}^{M-2,M}}\right), \nonumber
\end{eqnarray}
To first-order in perturbation theory these states are split
according to the diagonalization of the $3 \times 3$ matrix
\begin{widetext}
\begin{displaymath}
\left[\begin{array}{ccc}
V(M+1-\frac{4}{M}) + U(M-\frac{4}{M}) & U\sqrt{2(1-\frac{1}{M}-\frac{6}{M^2})} &  0 \\
U\sqrt{2(1-\frac{1}{M}-\frac{6}{M^2})} & V(M-1+\frac{2}{M}) + U(M-\frac{3}{M}) & U\sqrt{2(1-\frac{1}{M}-\frac{2}{M^2})} \\
 0 & U\sqrt{2(1-\frac{1}{M}-\frac{2}{M^2})} & V(M+1+\frac{5}{M}) + U(M-\frac{4}{M})
\end{array}\right].
\end{displaymath}
\end{widetext}
resulting in the non-degenerate excitations $\ket{\Psi_{{\rm
ex},1}^{M-2}}$, $\ket{\Psi_{{\rm ex},2}^{M-2}}$ and
$\ket{\Psi_{{\rm ex},3}^{M-2}}$.
\section{SF states}
In the SF regime ($U=V=0$) the ground state in the subspace
$\mathcal{S}_{2n}$ is given by
\begin{equation}
\ket{e_0}=\frac{1}{\sqrt{2}}\left(\kets{\Psi_{\rm{sf}}^{2n,2M-2n}}+\kets{\Psi_{\rm{sf}}^{2M-2n,2n}}\right)
\end{equation}
The excited states $\ket{ e_l }$ are constructed by expelling
pairs of atoms of opposite momentum $\pm k$. We can denote them by
$\ket{e_l} \equiv \ket{n_0,..,n_K}$ where $n_k=\frac{1}{2}\langle
\tt n_k^a+\tt n_k^b\rangle$ and we omit the negative values
because $n_{-k}=n_{k}$. An efficient way of generating the lowest
eigenstates in ascending order in energy for large $M$ is to
assume a quadratic approximation ${E}_{ap}(k) \sim k^2$ to the
dispersion relation ${E}(k)=2J(1-\cos(2 \pi k/M))$ for the single
particle energy in a tight-binding lattice. We can construct the
eigenstates in ascending order of energy as
\begin{eqnarray}
n_0& =& N-\sum_{k=1}^{K} n_k  \nonumber   \\
n_1 &=& m-\sum_{k=2}^{K} k^2 n_k  \nonumber \\
n_k &=& 0,..., \lfloor \frac{m-\sum_{q=2}^{k-1}q^2n_q}{k^2}\rfloor
\hspace{4mm} k=2,..,K \nonumber
\end{eqnarray}
where $m$ is an integer that runs from $0$ to some maximum value
$M_m$. Due to the color symmetry, the excited states $\ket{e_l}$
show a degeneracy $f(l)$ (see for example the first excited
degenerate states Eq.
 (\ref{eq:exc})) which can be calculated numerically.

\end{document}